\documentclass[ aip, PoF,
amsmath,amssymb,
reprint,%
]{revtex4-2}

\usepackage{dsfont}
\usepackage{mathtext}
\usepackage[utf8]{inputenc}

\usepackage{mathtools}
\usepackage[usenames,dvipsnames]{xcolor}
\usepackage{amsmath}
\usepackage{amssymb,mathrsfs}
\usepackage{graphicx}
\usepackage{epstopdf}
\usepackage[colorlinks=true]{hyperref}

\usepackage{mathtext}
\usepackage{subcaption}

\usepackage{etoolbox} 
\usepackage{lipsum} 
\usepackage[capitalize]{cleveref}

\makeatletter
\appto{\appendix}{%
	\@ifstar{\def\theequation@prefix{A.}}%
	{}%
}
\makeatother

\newcommand{\la}{\lambda}
\newcommand{\ga}{\gamma}

\newcommand{\om}{\omega}

\newcommand{\br}{{\bf r}}
\newcommand{\prt}{\partial}
\newcommand{\bu}{{\bf u}}
\newcommand{\bk}{{\bf k}}

\newcommand {\ox} {\overline{x}}

\begin{document}

\title{Propagation of wave packets along large-scale background waves }

\author{D.~V.~Shaykin}
\affiliation{Institute of Spectroscopy, Russian Academy of Sciences, Troitsk, Moscow, 108840, Russia}
\affiliation{Moscow Institute of Physics and Technology, Institutsky lane 9, Dolgoprudny, Moscow region, 141700, Russia}

\author{A.~M.~Kamchatnov}
\email[Author to whom correspondence should be addressed.  Electronic mail: ]{kamchatnov@gmail.com.}
\affiliation{Institute of Spectroscopy, Russian Academy of Sciences, Troitsk, Moscow, 108840, Russia}
\affiliation{Moscow Institute of Physics and Technology, Institutsky lane 9, Dolgoprudny, Moscow region, 141700, Russia}

\begin{abstract}
We study propagation of high-frequency wave packets along a large-scale background wave which evolves according to
dispersionless hydrodynamic equations for two variables (fluid density and flow velocity). Influence of the
wave packet on evolution of the background wave is neglected,
so the large-scale evolution can be found independently of the wave packet's motion. At the same time, propagation
of the packet depends in essential way on the background wave and it can be considered in framework of geometric
optics approximation with the use of Hamilton equations for the carrier wave number and the mean coordinate of the packet.
We derive equations for the carrier wave number as a function of the parameters which describe the background wave.
When they are solved, the path of the packet can be found by simple integration of the Hamilton equation.
The theory is illustrated by its application to the problem of propagation of wave packets along expanding large-scale
wave which evolution is described by the shallow water equations. In particular, they correspond to the
dispersionless limit of the defocusing nonlinear Schr\"{o}dinger equation, and then the expanding wave can be considered
as an expanding cloud of the Bose-Einstein condensate. Reflection of wave packets from upstream flows and their
propagation along stationary flows are also discussed. The analytical solutions found for these particular cases agree
very well with exact numerical solution of the nonlinear Schr\"{o}dinger equation.
\end{abstract}




\maketitle

\section{Introduction}\label{sec1}

Propagation of high-frequency wave packets through non-uniform and non-stationary media
is a well-developed field of wave physics. In this theory, it is usually assumed that there
exist three different scales of length: the wavelength $\la=2\pi/k$, corresponding to the
modulus $k=|\bk|$ of the carrier wave vector $\bk$, is much smaller that the size $d$ of the packet,
and $d$ is much smaller than the characteristic length $l$ at which parameters of the
medium considerably change,
\begin{equation}\label{eq1}
  \la\ll d\ll l.
\end{equation}
The difference of scales $d\ll l$ allows one to introduce a ``mean'' coordinate $\br$
of the packet with accuracy $d$, and the carrier wave vector $\bk$ is defined with
accuracy $2\pi/d\ll2\pi/\la=k$. Thus, in this geometrical optics approximation the packet's
dynamics is described by two variables $\br$ and $\bk$ which obey the Hamilton equations
\begin{equation}\label{eq2}
  \frac{d\br}{dt}=\frac{\prt\om}{\prt\bk},\qquad
  \frac{d\bk}{dt}=-\frac{\prt\om}{\prt\br},
\end{equation}
where $\om=\om(\bk,\br,t)$ is the dispersion function for linear waves propagating through
the medium whose parameters have the values corresponding to the point $\br$ at the moment $t$.
From an elementary physics point of view, the first equation (\ref{eq2}) is just a definition
of the group velocity of a packet and the second equation (\ref{eq2}) describes ``refraction''
of waves and can be considered as a generalization of Snell's refraction law. Coincidence of
Eqs.~(\ref{eq2}) with the Hamilton equations of classical mechanics of point particles is the
essence of the optical-mechanical analogy which, on one hand, played so important role in
development of quantum mechanics and, on the other hand, allows one to use for solving wave
propagation problems the well-developed methods of analytical mechanics (see, e.g.,
\cite{synge-37,ko-90} and references therein).

In practice, the possibility of analytical treatment of Eqs.~(\ref{eq2}) depends on specification
of properties of the medium and, consequently, on the form of the dispersion relation $\om=\om(\bk,\br,t)$.
As has recently been noticed \cite{ceh-19}, one such a fruitful specification appears when non-uniformity
and time-dependence are caused by a large-scale wave propagating through the medium. If such a
large-scale wave is described by two variables $\rho(\br,t)$ (``density'' of the medium) and
$\bu(\br,t)$ (its ``flow velocity''), then the packet's frequency $\om$ depends on $\br$ and $t$
only via the variables $\rho$ and $\bu$: $\om=\om(\bk; \rho(\br,t), \bu(\br,t))$. The problem of
propagation of wave packets was solved in Ref.~\cite{ceh-19} for a particular case of the large-scale
background wave in the form of a rarefaction wave in a unidirectional small-amplitude (KdV)
approximation with the use of the small-amplitude limit of the Whitham modulation equations
\cite{whitham-65,whitham-74} instead of Hamilton's equations (\ref{eq2}). This approach was
generalized in Ref.~\cite{MU} to the case of general simple waves in one-dimensional geometry (1D)
when $\rho$ and $u$ are functionally related with each other so that $\om$ becomes a function
of only one medium's parameter, $\om=\om(k,u(x,t))$, where $u(x,t)$ obeys in the simple wave
case to the Hopf equation
\begin{equation}\label{eq3}
  u_t+V_0(u)u_x=0.
\end{equation}
Combining Eqs.~(\ref{eq2}) and (\ref{eq3}), one readily gets~\cite{MU} the differential equation
for the function $k=k(u)$,
\begin{equation}\label{eq4}
\frac{dk}{du} = \frac{\prt \omega(k,u)/\prt u}{V_0(u)-\prt \omega(k,u)/\prt k}
\end{equation}
(it was first found in Ref.~\cite{el-05} as a consequence of Whitham's equations at the
small-amplitude edge of dispersive shock waves). Its solution $k=k(u)$ for a given initial
condition $k=k_0$ at $u=u_0$ together with the solution of the Hopf equation (\ref{eq3})
can be used for solving the first Hamilton equation (\ref{eq2}) (see Ref.~\cite{kamch-19a}),
so that one gets the path $x=x(t)$ of the packet along evolving background simple wave.

Although simple waves represent an important case of nonlinear waves appearing in fluid dynamics problems,
there are many situations described by general solutions of compressible fluid dynamics equations, when
there is no any functional relationship between $\rho$ and $u$. For example, such general solutions
appear in problems of expansion of gas clouds, problems of reflection of nonlinear waves from solid
walls, etc. (see, e.g., Ref.~\cite{LL6}). In this case, the wave number $k$ is affected by
two phenomena---refraction due to change of density and Doppler shift due to change of flow velocity,
which are independent of each other on the contrary to the simple wave case. The resulting value
$k=k(x,t)$ is to be obtained by means of integration of these changes along the packet's path which
itself depends on the carrier wave number $k$
as well as on the initial conditions $\rho=\rho_0(x),u=u_0(x)$ for the background flow. We will look for
situations when the wave number $k$ is the same function of the local values of $\rho$ and $u$,
$k=k(\rho(x,t),u(x,t))$, for any choice of
the initial conditions for the background flow. This greatly simplifies the analysis but imposes essential
conditions on the form of the function $k=k(\rho,u)$. Nevertheless, the inequalities (\ref{eq1})
allow one to distinguish situations when such a function $k=k(\rho,u)$ does exist in
the most physically important limit of high-frequency wave packets, i.e. in the limit of large $k$.

The aim of this paper is to develop the theory of propagation of wave packets along large-scale
background waves represented by general solutions of fluid dynamics equations. We will formulate
the applicability condition of our approach and will show in a particular case of generalized nonlinear
Schr\"{o}dinger (gNLS) equation that in the limit of large wave numbers the function $k=k(\rho,u)$
is given by a simple analytical expression. The theory is illustrated by several examples of large-scale
background flows and nonlinearity types and our analytical results agree very well with exact numerical
solutions of the gNLS equation for appropriate initial conditions.

\section{General theory}\label{sec2}

As was noticed in Ref.~\cite{sg-69}, dynamics of many physical systems depending on two parameters
can be written in hydrodynamics-like form
\begin{equation}\label{eq5}
\rho_t + (\rho u)_x = 0, \qquad (\rho u)_t + (\rho u^2 + P)_x = 0,
\end{equation}
where $\rho$ and $u$ are considered as `density' and `flow velocity' of the `fluid'. Consequently,
Eqs.~(\ref{eq5}) can be treated as conservation laws of mass and momentum, respectively. The function
$P=P(\rho,u,\rho_x,u_x,...)$ describes the action of pressure as well as effects of dispersion and/or
viscosity which contribution into the momentum density depends on gradients and higher order
derivatives of $\rho$ and $u$. Equations (\ref{eq5}) have uniform solutions $\rho=\overline{\rho}=\mathrm{const}$,
$u=\overline{u}=\mathrm{const}$. If we linearize these equations with respect to small deviations
$\rho'=\rho-\overline{\rho}$, $u'=u-\overline{u}$ and look for harmonic wave solutions
$\rho',u'\propto\exp[i(kx-\om t)]$, then we get the dispersion relation for linear waves
\begin{equation}\label{eq6}
  \om=\om(k,\overline{\rho},\overline{u}).
\end{equation}
It is worth noticing that usually one obtains two such expressions corresponding to linear waves which
propagate upstream or downstream the flow $\overline{u}$, but we do not distinguish them here in our notation.
In the theory of propagation of high-frequency wave packets the variables $\overline{\rho}$ and $\overline{u}$
can be replaced by the local background values $\rho$ and $u$ as long as they change slowly compared with
the packet's size (see Introduction). Of course, this long wave dynamics obeys the nonlinear
equations obtained from Eqs.~(\ref{eq5}) in dispersionless limit.

Equations for evolution of the background variables are obtained from Eqs.~(\ref{eq5}) when we neglect
the terms with derivatives in the function $P$ and assume that the limiting equation of state has the
form $P=P(\rho)$, so that the resulting equations can be written in the form
\begin{equation}\label{eq7}
\rho_t + (\rho u)_x = 0, \qquad u_t + uu_x+\frac{c^2}{\rho}\rho_x = 0,
\end{equation}
where $c=c(\rho)$ denotes the sound velocity ($c^2=dP/d\rho$), that is in the long wavelength limit
$k\to0$ the dispersion relation (\ref{eq6}) takes one of the forms
\begin{equation}\label{eq8}
  \frac{\om}{k}=v_+=u+c\quad\text{or}\quad \frac{\om}{k}=v_-=u-c
\end{equation}
for sound waves propagating upstream or downstream the flow $u$. The sound velocity $c=c(\rho)$ depends on
the local value of the density $\rho$ and can also serve as a local wave variable instead of $\rho$.

In the linear limit the variables $u'$ and $\rho'$ are related by the formulas $u'=\pm(c/\rho)\rho'$ for these
two `right' and `left' propagating waves with velocities $v_+$ and $v_-$, respectively. In a large scale
wave the variables $u'$ and $\rho'$ can be considered as differentials along small segments of such a wave,
and integration of the equation $du=\pm(c/\rho)d\rho$ yields two important variables
\begin{equation}\label{eq9}
  r_{\pm}=\frac{u}{2}\pm\frac12\int_0^{\rho}\frac{cd\rho}{\rho}
\end{equation}
called Riemann invariants (factor 1/2 is introduced for further convenience). If we invert the function
$c=c(\rho)$ and substitute $\rho=\rho(c)$ into Eq.~(\ref{eq9}), then we get the Riemann invariants in the
form
\begin{equation}\label{eq10}
  r_{\pm}=\frac{u}{2}\pm\sigma(c),\qquad \sigma(c)=\frac12\int_0^c\frac{c}{\rho(c)}\frac{d\rho}{dc}dc.
\end{equation}
Evidently, all physical variables $u,\rho,c$ can be expressed in terms of the Riemann invariants,
\begin{equation}\label{eq11}
  u=r_++r_-,\quad \rho=\rho(r_+,r_-),\quad c=c(r_+,r_-).
\end{equation}
Equations (\ref{eq7}) written in terms of $r_{\pm}$ take very simple diagonal form
\begin{equation}\label{eq12}
\frac{\prt r_+ }{\prt t} + v_+ \frac{\prt r_+}{\prt x}=0,\quad
\frac{\prt r_- }{\prt t} + v_- \frac{\prt r_-}{\prt x}=0,
\end{equation}
where the  characteristic velocities are equal to
\begin{equation}\label{eq13}
v_{\pm}=v_{\pm}(r_+,r_-) = r_++r_-\pm c(r_+,r_-).
\end{equation}
Simple waves correspond to unidirectional flows with one of the Riemann invariants constant. Then one of
Eqs.~(\ref{eq12}) is satisfied identically and the other one reduces to the Hopf equation, so we return
to situations considered in Ref.~\cite{MU}. In general solutions both Riemann invariants $r_{\pm}$
change with space coordinate $x$ and time $t$, so they can serve as new coordinates in the region of the
general solution. This hodograph transform makes $x$ and $t$ functions of $r_+$ and $r_-$,
\begin{equation}\label{eq14}
x = x(r_+,r_-), \qquad t = t(r_+,r_-),
\end{equation}
and it casts Eqs.~(\ref{eq12}) to linear equations for $x$ and $t$,
\begin{equation}\label{eq15}
  \frac{\prt x}{\prt r_+}-v_-\frac{\prt t}{\prt r_+}=0,\quad
  \frac{\prt x}{\prt r_-}-v_+\frac{\prt t}{\prt r_-}=0,
\end{equation}
which can be solved, for example, by the hodograph method of Ref.~\cite{tsarev}. If the solution
(\ref{eq14}) is known, then either it can be inverted to give explicit formulas for the Riemann invariants
\begin{equation}\label{eq16}
r_+ = r_+(x,t), \qquad r_- = r_-(x,t)
\end{equation}
and, consequently, for the physical variables (\ref{eq11}), or it gives these physical variables in
a parametric form.

When a high-frequency wave packet propagates along the background wave given by the general solution of
hydrodynamic equations in the Riemann diagonal form (\ref{eq12}), then both the dispersion relation and the
group velocity become functions of the Riemann invariants,
\begin{equation}\label{eq17}
  \om=\om(k,r_+,r_-),\qquad v_g=\frac{\prt\om}{\prt k}=v_g(k,r_+,r_-).
\end{equation}
Let the packet enter into the region of the general solution at some point $x_0=x(r_+^0,r_-^0)$ at the
moment $t_0=t(r_+^0,r_-^0)$ with the carrier wave number $k_0$. Then its values at later moments of time
depend on the path in the $(r_+,r_-)$ plane from the initial point $(r_+^0,r_-^0)$ to the final point
$(r_+,r_-)$. Motion of the packet is governed by the 1D version of Hamilton's equations (\ref{eq2}),
\begin{equation}\label{eq19}
\frac{d x}{d t} = \frac{\prt \omega }{\prt k},\qquad
\frac{d k}{d t} = -\frac{\prt \omega }{\prt x}.
\end{equation}
Then the second equation gives
$$
\frac{dk}{dt}=-\frac{\prt \omega }{\prt r_+}\frac{\prt r_+ }{\prt x}
-\frac{\prt \omega }{\prt r_-}\frac{\prt r_- }{\prt x}.
$$
In Eqs.~(\ref{eq17}) the frequency $\om=\om(k,r_+,r_-)$ depends, besides $k$, only on local
values $r_{\pm}$ of the Riemann invariants. We assume here that the wave number $k$ is also a function
of the Riemann invariants, $k=k(r_+,r_-)$. Then along the packet's path we have
\begin{equation}\nonumber
  \begin{split}
  \frac{dk}{dt}&=\frac{\prt k }{\prt r_+}\frac{d r_+ }{dt}+\frac{\prt k }{\prt r_-}\frac{d r_- }{dt}\\
&=\frac{\prt k }{\prt r_+}\left(\frac{\prt r_+ }{\prt t}+v_g\frac{\prt r_+ }{\prt x}\right)
+\frac{\prt k }{\prt r_-}\left(\frac{\prt r_- }{\prt t}+v_g\frac{\prt r_- }{\prt x}\right) \\
&=-\frac{\prt k }{\prt r_+}(v_+-v_g)\frac{\prt r_+ }{\prt x}
-\frac{\prt k }{\prt r_-}(v_--v_g)\frac{\prt r_- }{\prt x},
  \end{split}
\end{equation}
where we have used Eqs.~(\ref{eq12}). Comparison of these two expressions shows that they are
consistent with each other, if the condition
\begin{equation}\label{eq19a}
\begin{split}
  &\left[\frac{\prt k }{\prt r_+}(v_+-v_g)-\frac{\prt \om }{\prt r_+}\right]\frac{\prt r_+}{\prt x}\\
&+\left[\frac{\prt k }{\prt r_-}(v_--v_g)-\frac{\prt \om }{\prt r_-}\right]\frac{\prt r_-}{\prt x} =0
\end{split}
\end{equation}
is fulfilled. As we assumed above, the function $k=k(x,t)$ reduces in situations under
consideration to the function $k=k(r_+(x,t),r_-(x,t))$ independent of the initial conditions,
so the expressions in square brackets are only
functions of $r_+$ and $r_-$. Let us denote them as $X_+(r_+,r_-)$ and $X_-(r_+,r_-)$. At the same
time $r_+(x,t)$ and $r_-(x,t)$ depend on the initial conditions $r_+^{0}(x),r_-^{0}(x)$, so the
derivatives $\prt r_{\pm}/\prt x$ at the points with the same values of $r_{\pm}$ can be different
for different choices of the initial conditions. Let us choose two such initial conditions
$r_{\pm}^{0(1)}(x),r_{\pm}^{0(2)}(x)$, that the corresponding solutions $r_{\pm}^{(1)}(x,t),r_{\pm}^{(2)}(x,t)$
of Eqs.~(\ref{eq12}) have at the points with $r_{\pm}^{(1)}=r_{\pm}^{(2)}\equiv r_{\pm}$ the
derivatives with respect to $x$ satisfying the inequality
$$
\frac{\prt r_+^{(1)}}{\prt x}\cdot\frac{\prt r_-^{(2)}}{\prt x}-
\frac{\prt r_-^{(1)}}{\prt x}\cdot\frac{\prt r_+^{(2)}}{\prt x}\neq 0.
$$
Then the system
\begin{equation}\nonumber
\begin{split}
  &X_+(r_+,r_-)\frac{\prt r_+^{(1)}}{\prt x}+X_-(r_+,r_-)\frac{\prt r_-^{(1)}}{\prt x}=0,\\
&X_+(r_+,r_-)\frac{\prt r_+^{(2)}}{\prt x}+X_-(r_+,r_-)\frac{\prt r_-^{(2)}}{\prt x}=0
\end{split}
\end{equation}
gives $X_+(r_+,r_-)=0,X_-(r_+,r_-)=0$, and the function
$k=k(r_+,r_-)$ must satisfy the equations
\begin{equation}\label{eq20}
\frac{\prt k}{\prt r_+} = \frac{\prt \omega / \prt r_+}{v_+-v_g}, \qquad
\frac{\prt k}{\prt r_-} = \frac{\prt \omega / \prt r_-}{v_--v_g}.
\end{equation}
For existence of such a function $k=k(r_+,r_-)$ these derivatives must commute
\begin{equation}\label{eq20a}
  \frac{\prt}{\prt r_+}\left(\frac{\prt k}{\prt r_-}\right)=
  \frac{\prt}{\prt r_-}\left(\frac{\prt k}{\prt r_+}\right).
\end{equation}
If Eqs.~(\ref{eq20}) satisfy this condition, then we look for their solution $k=k(r_+,r_-,q)$,
where $q$ is an integration constant which value is determined by the initial value $k_0$ of
the wave number at the point $x_0$ and moment of time $t_0$ with local values of the Riemann
invariants $r_{\pm}^0=r_{\pm}(x_0,t_0)$. As a result, the solution $k=k(r_+,r_-,q)$ defines the
carrier wave number $k$ in the whole region of the solution $r_{\pm}=r_{\pm}(x,t)$ of
hydrodynamic equations (\ref{eq12}).
Substitution of these functions into the first
Hamilton equation (\ref{eq19}) gives the equation
\begin{equation}\label{eq21}
\frac{dx}{dt} = \left.\frac{\prt}{\prt k} \omega
\big(k, r_+(x,t),r_-(x,t) \big)\right|_{k=k[r_+(x,t),r_-(x,t)]}
\end{equation}
for the packet's path $x=x(t)$.

If the condition (\ref{eq20a}) is not fulfilled exactly, then we should confine ourselves
to an approximate solution correct in the limit of large $k$ and such a solution would be enough for our
treatment of propagation of high-frequency wave packets. To formulate the conditions of applicability
of such an approximation, we assume that the flow and sound velocities have the same order of magnitude,
$|u|\sim c$, and therefore $r_{\pm}\sim c$. Now we look for an approximate solution for $k\gg c$, where
$k\sim k_0$, $k_0$ being the initial value of the wave number. If we denote the right-hand sides of
Eqs.~(\ref{eq20}) as $R_{\pm}(k,r_+,r_-)$, then we expand them in series with respect to small
parameters $|r_{\pm}|/k$ and keep only the terms $R_{\pm}^{\text{asymp}}(k,r_+,r_-)$ satisfying
the commutativity condition (\ref{eq20a}). The solution $k=k^{\text{asymp}}(r_+,r_-,q)$ of the
resulting equations
\begin{equation}\label{eq20b}
  \frac{\prt k}{\prt r_{\pm}}=R_{\pm}^{\text{asymp}}(k,r_+,r_-)
\end{equation}
should also be looked for with the same accuracy.

If the solution of hydrodynamic equations (\ref{eq7}) for the background wave is found directly for
the physical variables $\rho$ and $u$, then it is convenient to write equations for $k=k(\rho,u)$
without transition to the Riemann invariants although they may be known. Simple calculation
similar to derivation of Eqs.~(\ref{eq20}) gives
\begin{equation}\label{eq25}
\begin{split}
&\frac{\prt k}{\prt \rho} = \frac{(v_g-u)\frac{\prt \omega}{\prt \rho} +
\frac{c^2}{\rho}\frac{\prt \omega}{\prt  u}}{c^2-(v_g - u)^2},\\
&\frac{\prt k}{\prt u} = \frac{(v_g-u)\frac{\prt \omega}{\prt u} +
\rho\frac{\prt \omega}{\prt  \rho}}{c^2-(v_g - u)^2}.
\end{split}
\end{equation}
Naturally, these derivatives should satisfy the commutativity condition
$\frac{\prt}{\prt u}\left(\frac{\prt k}{\prt \rho}\right)=
\frac{\prt}{\prt \rho}\left(\frac{\prt k}{\prt u}\right)$
at least in the limit of large $k$, when the asymptotic solution $k=k^{\text{asymp}}(\rho,u,q)$
can be found.

Let us illustrate the general theory by its application to systems whose evolution
is governed by the gNLS equation.

\section{Application to gNLS equation}\label{sec3}

Here we shall apply the developed above approach to the gNLS equation
\begin{equation}\label{eq26}
i\psi_t+\frac{1}{2}\psi_{xx} - f(|\psi|^2)\psi = 0,
\end{equation}
which has a number of physical applications and is written here in standard
dimensionless variables. To be definite, we shall imply here that this equation
describes dynamics of Bose-Einstein condensates (BECs) with different kinds of repulsive
interaction between atoms (or quasiparticles in case of polariton condensates). To cast this
equation to the form (\ref{eq5}), we make the Madelung transformation of the BEC's wave function
\begin{equation}\nonumber
\psi(x,t) = \sqrt{\rho(x,t)}\exp\left(i\int^x u(q,t)dq\right)
\end{equation}
and obtain the system
\begin{equation}\label{eq28}
\begin{split}
&\rho_t + (\rho u)_x = 0, \\
&u_t + uu_x + \frac{c^2}{\rho} \rho_x + \bigg( \frac{\rho_x^2}{8\rho^2}
- \frac{\rho_{xx}}{4\rho} \bigg)_x = 0,
\end{split}
\end{equation}
where
\begin{equation}\label{eq28b}
  c^2=\rho f'(\rho)
\end{equation}
is a function of the condensate's density $\rho=|\psi|^2$ and $u$ has a meaning of its flow velocity.
The last term in the second equation (\ref{eq28}) is responsible for the dispersive effects, so
linearization of these equations with respect to small deviations from a uniform flow with constant
density and flow velocity yields the Bogoliubov dispersion relation
\begin{equation}\label{eq29}
\omega = k\left( u \pm \sqrt{c^2+\frac{k^2}{4}} \right).
\end{equation}
In the limit $k\to0$ we obtain the characteristic velocities (\ref{eq8}), consequently $c$
equals to the sound velocity of long waves.
Dynamics of a large scale wave obeys the dispersionless `shallow water' equations
(\ref{eq7}) which are obtained from (\ref{eq28}) by neglecting the dispersion terms.

Formula (\ref{eq28b}) determines $c$ as a function of $\rho$, $c=c(\rho)$. It is convenient to
invert this function and then the function $\rho=\rho(c)$ characterizes the nonlinear properties of
the system under consideration. Thus, in this notation the large scale flow is described by the
variables $c(x,t),u(x,t)$ which are equivalent to the previous variables $\rho(x,t), u(x,t)$ but
they are more convenient in some calculations. Then we easily obtain
\begin{equation}\nonumber
\begin{split}
  &v_g=\frac{\prt\om}{\prt k}=u+\frac{k^2+2c^2}{\sqrt{k^2+4c^2}},\qquad \frac{\prt\om}{\prt u}=k,\\
 & \frac{\prt\om}{\prt \rho}=\frac{2ck/\rho'}{\sqrt{k^2+4c^2}},
  \end{split}
\end{equation}
and substitution of these formulas into Eqs.~(\ref{eq25}) casts them to the form
\begin{equation}\label{eq30}
  \begin{split}
  & \frac{\prt k}{\prt c}=-\frac{c[(2+c\rho'/\rho)k^2+4(1+c\rho'/\rho)c^2]}{k(k^2+3c^2)},\\
  & \frac{\prt k}{\prt u}=-\frac{\sqrt{k^2+4c^2}[k^2+2(1+\rho/(c\rho'))c^2]}{k(k^2+3c^2)}.
  \end{split}
\end{equation}
The difference of cross-derivatives equals to
\begin{equation}\label{eq31}
  \begin{split}
  &\frac{\prt}{\prt c}\left(\frac{\prt k}{\prt u}\right)-\frac{\prt}{\prt u}\left(\frac{\prt k}{\prt c}\right)
  =\frac{\sqrt{k^2+4c^2}}{k\rho\rho'(k^2+3c^2)^2}\\
  &\times\left[(k^2+6c^2)\rho^{\prime 2}(\rho'-2c)+2(k^2+3c^2)\rho^2(c\rho^{\prime\prime}-\rho')\right].
  \end{split}
\end{equation}
As one can see, this difference only vanishes in the case of $\rho=c^2$ which corresponds to the
standard NLS equation (\ref{eq26}) with $f(\rho)=\rho$. In the limit of large $k$ this difference
vanishes as $\propto k^{-2}$. Thus, in the limit of large $k$ the condition (\ref{eq20a}) is fulfilled
and the wave number $k$  becomes a
function of two variables, $k=k(c,u)$. In this limit the corresponding partial derivatives are given by the
asymptotic expressions
\begin{equation}\label{eq31c}
  \frac{\prt k^2}{\prt c^2}=-2\left(1+\frac{c^2}{\rho}\frac{d\rho}{d c^2}\right),\qquad
  \frac{\prt k}{\prt u}=-1,
\end{equation}
obtained from Eqs.~(\ref{eq30}) for large $k$. To solve these equations in the asymptotic
limit, we notice that the second equation gives $k=q-u+F(c^2)$, where in the main approximation
$k\approx q\sim k_0\gg|u|,c$. We suppose that $F(c^2)\sim c^2$ (this will be confirmed by the final result),
so $k^2\approx(q-u)^2+2qF(c^2)$, where we have neglected small terms $uF\sim c^3$ and $F^2\sim c^4$.
Then the first equation (\ref{eq31c}) gives at once
$$
F(c^2)=-\frac1q\left(c^2+\int_0^{\rho(c)}\frac{c^2}{\rho}d\rho\right)\sim\frac{c^2}{q}
$$
and we arrive at the asymptotic solution
\begin{equation}\label{eq31d}
  k^2=(q-u)^2-2\left(c^2+\int_0^{\rho(c)}\frac{c^2}{\rho}d\rho\right).
\end{equation}
Here $q$ is an integration constant determined by the value of $k=k_0$ at some initial point in the $(c,u)$-plane
and, hence, $q=k_0\gg|u|,c$ in this asymptotic solution.
For an important case of the nonlinearity function
\begin{equation}\label{eq31e}
  f(\rho)=\frac1{\ga-1}\rho^{\ga-1},\quad\text{when}\quad c^2=\rho^{\ga-1},\quad \rho=c^{2/(\ga-1)},
\end{equation}
we get
\begin{equation}\label{eq31f}
  k^2=(q-u)^2-\frac{2\ga}{\ga-1}c^2,
\end{equation}
where $\ga$ is an effective `adiabatic constant' (in gNLS equation case it does not have a meaning of ratio
of specific heat capacities, of course). If $\ga=2$, then the formula
\begin{equation}\label{eq31g}
  k^2=(q-u)^2-4c^2
\end{equation}
gives the exact solution of Eqs.~(\ref{eq30}) with $\rho=c^2$, when these derivatives commute.
Expression (\ref{eq31d}) for the function $k=k(c,u)$ and its particular cases (\ref{eq31f}) and
(\ref{eq31g}) can be used for solving various problems on propagation of high-frequency
wave packets.

\section{Propagation along expanding BEC}

\subsection{NLS equation $(\ga=2)$}

At first we shall consider the case of the standard NLS equation with the nonlinearity function
$f(\rho)=\rho$, $(\ga=2)$.
Let at the initial moment of time the BEC be confined in a standard trap with harmonic potential,
so in the Thomas-Fermi approximation distribution of the density has a form of inverted parabola, and
$u$ is equal to zero everywhere
\begin{equation}\label{eq32}
\begin{cases}
\rho(x,0) = a^2\big[ 1-\big(\frac{x}{l}\big)^2\big],\quad |x|\leq l,\\
u(x,0)=0,
\end{cases}
\end{equation}
where $l\gg1$ is the size of the BEC cloud and $a^2$ is the maximal density at the center $x=0$
of the trap. At the moment $t=0$ the trap is switched off and the condensate starts its expansion.
Equations (\ref{eq7}) with the initial conditions (\ref{eq32}) were solved in
Ref.~\cite{bkk-03} with the use of {\it ansatz} suggested in Ref.~\cite{talanov-65} for description
of self-focusing of light beams propagating through nonlinear medium. The exact solution has the form
\begin{equation}\label{eq33}
\begin{cases}
\rho(x,t) = \frac{a^2}{f(t)}\big[ 1-\big(\frac{x}{lf(t)}\big)^2\big],\quad |x|\leq lf(t),\\
u(x,t)=\phi(t) x,
\end{cases}
\end{equation}
where $f(0)=1, \phi(0) = 0$ according to the initial conditions (\ref{eq32}). Substitution of these
expressions into \eqref{eq7} leads to the system for the functions $f(t), \phi (t)$:
\begin{equation}\label{eq34}
\phi = \frac{1}{f}\frac{df}{dt}, \qquad \frac{d^2f}{dt^2} = \frac{2a^2}{l^2 }\frac1{f^2},
\end{equation}
These equations can be readily solved to give the function $f(t)$ in an implicit form
\begin{equation}\label{eq35}
\frac{2a}{l} t= \sqrt{f(f-1)} +\ln\left(\sqrt{f-1}+\sqrt{f}\right),
\end{equation}
and then the first expression in Eqs.~(\ref{eq34}) defines the function $\phi(t)$.
Equations \eqref{eq33}-\eqref{eq35} describe in a parametric form the large scale evolution of the
condensate cloud.

\begin{figure}[t]
\begin{center}
	\begin{subfigure}{3.7cm}	
	\caption{}
	\includegraphics[width = 3.7cm,height = 3.7cm]{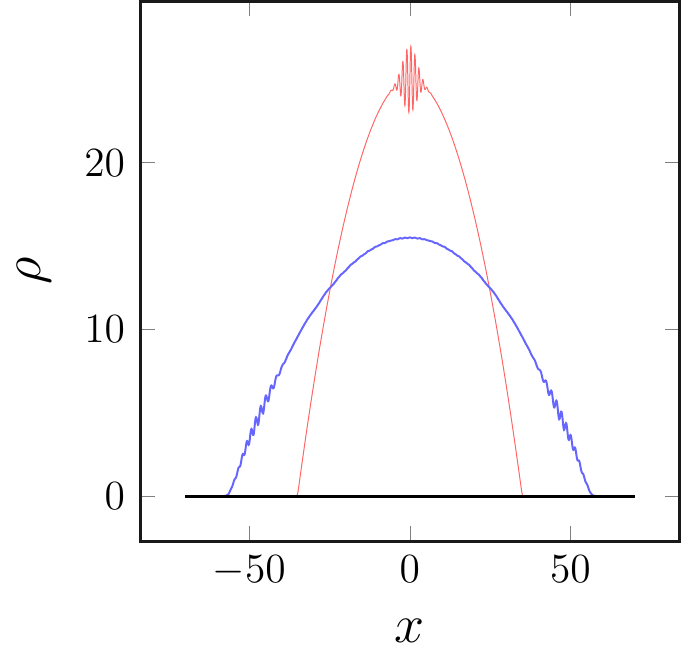}
	\end{subfigure}
\hspace{2mm}
	\begin{subfigure}{3.7cm}
	\caption{}
	\includegraphics[width = 3.7cm,height = 3.7cm]{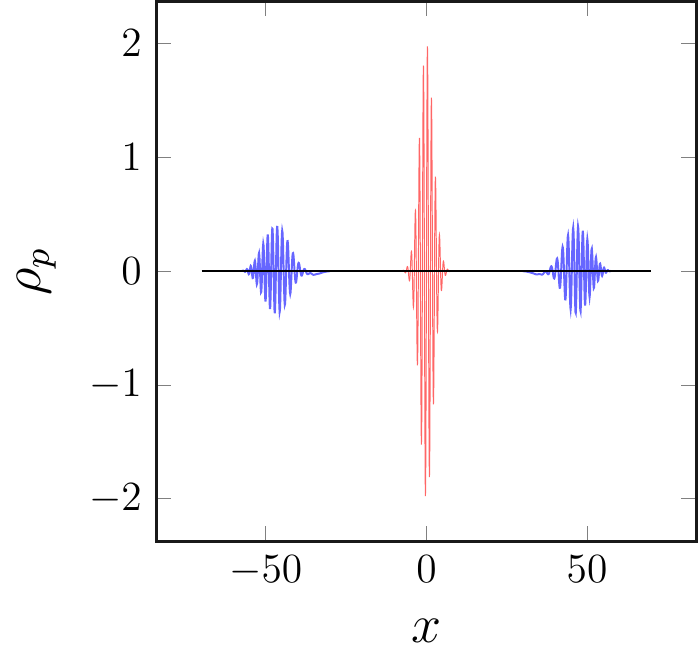}
	\end{subfigure}
\caption{(a) Profiles $\rho(x,t)$ of the density obtained by numerical solution of
the Gross-Pitaevskii equation with the initial distribution \eqref{eq32} for
$a = 5$, $l = 35$, $k_0=5$ and the wave packet added at the center; the curves correspond
to the moments of time $t=0$ (red), $t=35$ (blue).
 (b) Profiles of density in the linear wave packet obtained by subtraction of
the result of evolution of the smooth background pulse from the evolution of the whole
condensate with account of contribution of the wave packet at the same moments of time.
}
\label{fig1}
\end{center}
\end{figure}

\begin{figure}[t]
\begin{center}
	\includegraphics[width = 6.5cm,height = 6.5cm]{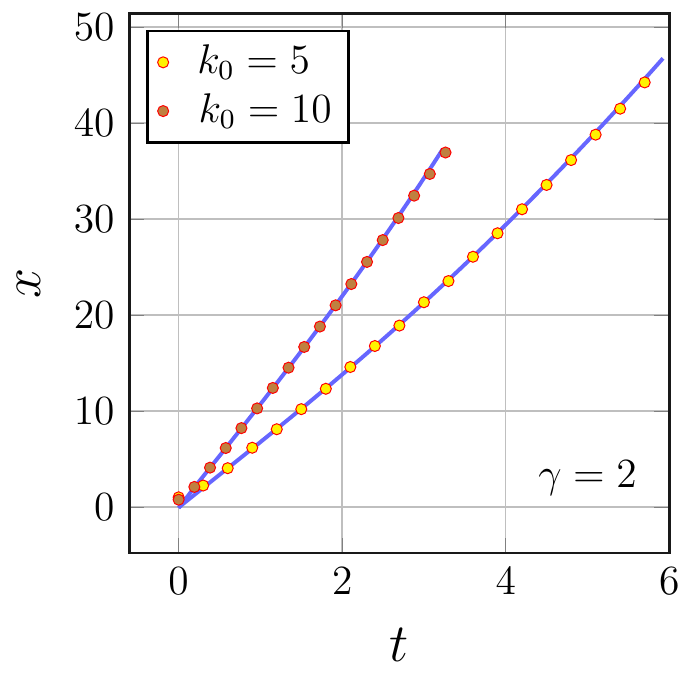}
\caption{Paths $x(t)$ of the wave packets along expanding BEC $(\ga=2)$ for two values of the
initial carrier wave vector $k_0=5$ or $k_0=10$.
 }
\label{fig2}
\end{center}
\end{figure}

Now we assume that a high-frequency disturbance of condensate's density is located in vicinity of
its center at the moment $t=0$ of switching off the trap. This disturbance can be represented as
a wave packet made from the Bogoliubov harmonics with both signs in Eq.~(\ref{eq29}) and some carrier
wave vector $k_0$. Consequently, this disturbance splits to two wave packets propagating symmetrically
along the expanding cloud of BEC (see Fig.~\ref{fig1}) and the carrier wave vector evolves according
to the equations derived in Section~\ref{sec3}. As was noticed above, in this case Eqs.~(\ref{eq30})
take simple form
\begin{equation}\label{eq36}
\begin{split}
\frac{\prt k}{\prt c} = -4\frac{c}{k},\qquad
\frac{\prt k}{\prt u} = -\frac{\sqrt{k^2+4c^2}}{k},
\end{split}
\end{equation}
and these derivatives commute with each other. It is remarkable that this case
corresponds to the complete integrability of the NLS equation, so one may suppose that these two
properties can be related with each other. The exact solution of these equations coincides with the
solution (\ref{eq31g}) of their large $k$ approximations.
Equation (\ref{eq21}) for the packet's path takes with account of Eqs.~(\ref{eq29}), (\ref{eq31g})
the form (we assume here $q-u > 0$)
\begin{equation}\label{eq43.1}
\frac{dx}{dt} = q - \frac{2\rho(x,t)}{q - u(x,t)},
\end{equation}
and it can be easily solved numerically for the given distributions (\ref{eq33})-(\ref{eq35}) of
$\rho(x,t)$ and $u(x,t)$. We compared this theory with the exact numerical solution of the
Gross-Pitaevskii equation. To this end, we have chosen the initial distributions \eqref{eq32}
with the parameters $a = 5$, $l = 35$ and added at the center $x = 0$ a small perturbation in
the form of the wave packet with $k_0 = 5$ or $k_0=10$. In numerical solution of Eq.~\eqref{eq26}
the initial disturbance splits into two packets (see Fig.~\ref{fig1}) and we will consider
the right-propagating packet. Since splitting of the initial wave packet to the
right and left propagating packets takes some time, their coordinates may be not well enough
defined at small values of time and in these cases the numerical corresponding points are not
depicted in our figures. The paths of the wave packet for these two initial
values of $k_0$ are shown in Fig.~\ref{fig2},
where the solid line depicts the solution of Eq.~(\ref{eq43.1}) and the dots correspond to
locations of the packet extracted from the full numerical solution of the Gross-Pitaevskii
equation (\ref{eq26}). As one can see, our analytical approach
agrees very well with the exact numerical solution even for relatively small values of $k$.

\subsection{gNLS equation $(\ga=3)$}

Now we shall consider the case of the gNLS equation (\ref{eq26}) with $f(\rho)=\rho^2/2$, so that
$c=\rho$ (see Eq.~(\ref{eq31e})). In this case it is convenient to redefine the Riemann
invariants (\ref{eq10}) by multiplying them by 2 so we get $r_{\pm}=v_{\pm}=u\pm c$ and
\begin{equation}\label{eq37}
   u=\frac12(r_++r_-),\quad c=\rho=\frac12(r_+-r_-).
\end{equation}
As a result, the hydrodynamic equations (\ref{eq12}) for the background flow split into two
independent Hopf equations
\begin{equation}\label{eq38}
  \frac{\prt r_{\pm}}{\prt t}+r_{\pm} \frac{\prt r_{\pm}}{\prt x}=0.
\end{equation}
Such a splitting to two separate Hopf equations can be related with a
specific symmetry of the gNLS equation for the case $\ga=3$. This symmetry was first found
in Ref.~\cite{talanov-70} for NLS with two spatial dimensions and generalized in Ref.~\cite{kt-85}
to gNLS with any number of spatial dimensions so that $\ga=3$ corresponds to our case with one
spatial dimension.
If in the initial state the flow velocity equals everywhere to zero, then the initial
distributions of the Riemann invariants only differ by signs, $r_+(x,0)=c(x,0)$,  $r_-(x,0)=-c(x,0)$,
so the solutions of the Hopf equations are given by the formulas
\begin{equation}\label{eq39}
  x-r_+t=\ox(r_+),\qquad x-r_-t=\ox(-r_-),
\end{equation}
where $\ox(r_+)$ is the function inverse to $r_+=r_+(x,0)$.

\begin{figure}[t]
\begin{center}
	\includegraphics[width = 6.5cm,height = 6.5cm]{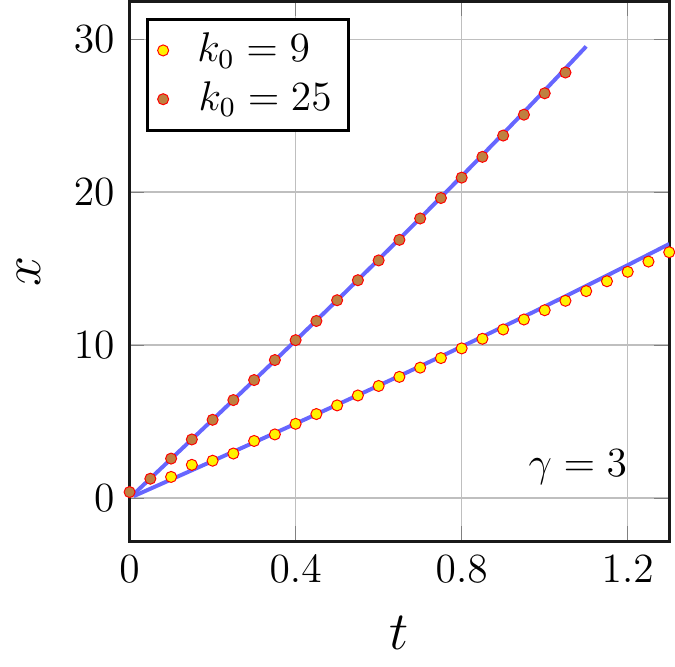}
\caption{Paths $x(t)$ of the wave packets along expanding BEC $(\ga=3)$ for two values of the
initial carrier wave vector $k_0=9$ or $k_0=25$.
 }
\label{fig3}
\end{center}
\end{figure}

For simplicity we take again the initial distribution of $\rho=c$ in a parabolic form (\ref{eq32}),
so that $\ox(r)=l\sqrt{1-r/a^2}$. Then the solution (\ref{eq39}) can be written as
\begin{equation}\label{eq40}
\begin{split}
  &x-(u+c)t=l\sqrt{1-(u+c)/a^2},\\
  & x-(u-c)t=l\sqrt{1+(u-c)/a^2},
  \end{split}
\end{equation}
and the time-dependent distributions $c=c(x,t), u=u(x,t)$ can be easily expressed by relatively simple
analytical formulas
\begin{equation}\label{eq41}
\begin{split}
  \rho(x,t) &= c(x,t) = \frac{l}{4a^2t^2}  \sqrt{l^2+4a^4l^2 +4a^2tx}  +\\
  &+ \frac{l}{4a^2t^2} \big( -2l +\sqrt{l^2+4a^4l^2 -4a^2tx} \big),
  \end{split}
\end{equation}
\begin{equation}\label{eq42}
\begin{split}
  u(x,t) &= \frac{x}{t} + \frac{l}{4a^2t^2}\sqrt{l^2+4a^4l^2 -4a^2tx} - \\
  & - \frac{l}{4a^2t^2}  \sqrt{l^2+4a^4l^2 +4a^2tx}
  \end{split}
\end{equation}
Equation (\ref{eq21}) takes (with account of $k^2=(q-u)^2-3c^2$, see Eq.~(\ref{eq31f})) the form
\begin{equation}\label{eq43.2}
  \frac{dx}{dt}=u+\frac{(q-u)^2-c^2}{\sqrt{(q-u)^2+c^2}},
\end{equation}
where $u=u(x,t)$ and $c=c(x,t)$ are known functions. Again we put a wave packet at the top of
the density distribution with some initial value of the carrier wave number $k_0$, which
determines the value of $q$, and solve Eq.~(\ref{eq43.2}) numerically. The results are compared with
the exact solution of the full gNLS equation, see Fig.~\ref{fig3}, and one can see that the
approximate theory using the asymptotic formula (\ref{eq31f}) agrees very well with the exact
solutions.

\section{Packet's reflection from an upstream flow}

So far we have considered propagation of a packet along a single background wave with
coinciding directions of the background flow and the packet's propagation. It is of considerable
interest to discuss situations when the packet with the fixed value of $q$ propagates
through the region of the background flow where the group velocity can
vanish. In the optical-mechanical analogy this corresponds to the `turning points' where
the packet changes direction of its propagation. Due to presence of the flow velocity this
does not mean that the wave vector vanishes, so the asymptotic theory can be still
applicable. In this section, we will check the validity
of Eq.~(\ref{eq31g}) ($\ga=2$) with the same value of $q$ after such a reflection of
a wave packet from the turning point.

\begin{figure}[t]
\begin{center}
	\includegraphics[width = 6.5cm,height = 6.5cm]{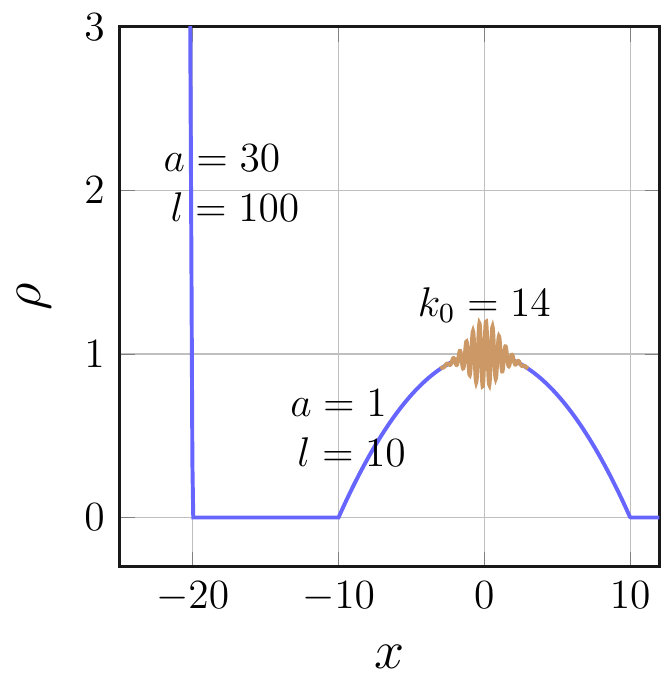}
\caption{Initial distributions of two BEC clouds in the form of \eqref{eq32} with parameters
indicated in the picture. The wave packet at $t=0$ is drawn in brown, it splits for $t>0$ to two modes,
one of which propagates to the left, the other to the right. Our aim is to investigate the left
moving packet with $q = -\sqrt{14^2+4}$.
 }
\label{fig4}
\end{center}
\end{figure}

\begin{figure}[t]
\begin{center}
	\includegraphics[width = 6.5cm,height = 6.6cm]{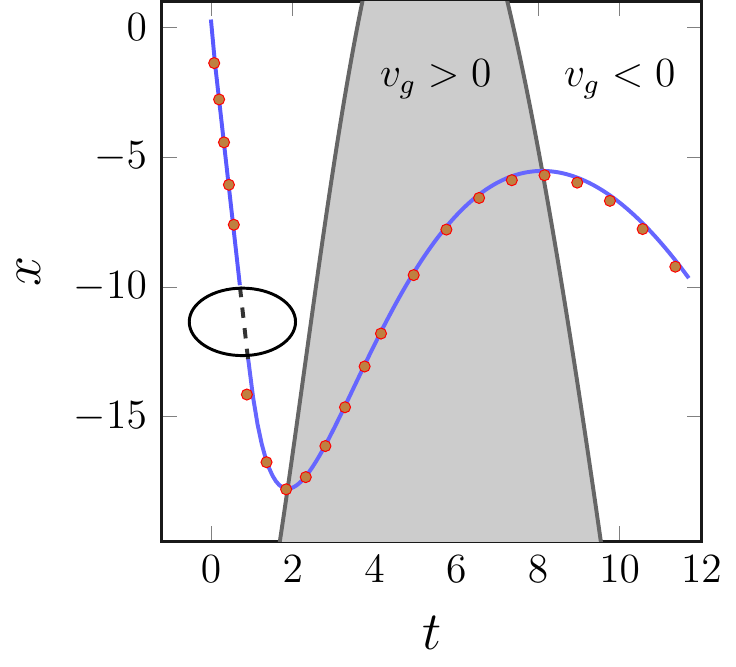}
	\caption{Comparison of the trajectories obtained by numerical solution of the NLS equation (circles)
and by solutions of Eq.~\ref{eq43.1}. The dashed line inside the ellipse shows the transition through the empty
region with group velocity $v_g=q$. In the grey and white areas the packets with the fixed value of $q=-\sqrt{14^2+4}$
have positive and negative velocities, respectively.
 }
\label{fig5}
\end{center}
\end{figure}

To this end, we launch a relatively slow packet towards a large expanding condensate in the following way.
At the initial moment of time we form two clouds with parabolic distributions of density: a large cloud
occupies the region with $l=100$, its right edge is located at $x=-20$, and the field amplitude equals to $a=30$,
a small cloud occupies the region $-10<x<10$, so its left edge is located at $x=-10$, and the amplitude
equals to $a=1$. The wave packet is generated initially at the top of the small cloud around the point $x=0$
with the wave number $k_0=14$, see Fig.~\ref{fig4}. Then the clouds start their expansions, the initial packet
splits to two packets, and the left packet with $q=-\sqrt{14^2+4}\approx -14.14$ propagates
first towards the empty region between two clouds, propagates through it, and reaches finally the large cloud
with upstream background flow. In this case the first Hamilton equation (with sign ``--'' in the dispersion
relation and $q-u<0$) has the same form as \eqref{eq43.1}. Interestingly enough, propagation through the empty region does not look
in numerical calculations like motion of a well-defined packet: it disperses at the left edge of the small cloud
into a wide distribution of small-amplitude waves which are collected again into a narrow packet at the
right edge of the large cloud.
The essential point is that at the edges of the both clouds, where $\rho=0$, we find from Eqs.~\eqref{eq43.1}
that the following estimate holds at the edges of the clouds:
\begin{equation}\label{v=q}
\left.v_g\right|_{\text{edges}}=  q.
\end{equation}
This allows us to find analytically trajectories of the packet along three characteristic regions: the small
cloud, the empty region, and the large cloud; they are shown in Fig.~\ref{fig5} by a solid line. Positions
of the packet at different moments of time are shown by dots; as was indicated above, they cannot be
identified within the empty region. The packet enters the big cloud with negative group velocity, at some point
the group velocity vanishes due to increase of the upstream background flow velocity, and later vanishes
again because the flow velocity is small
near the center of the cloud. The boundaries of the gray region are obtained by equating the right-hand side
of Eq.~(\ref{eq43.1}) to zero for given distributions of $\rho(x,t)$ and $u(x,t)$. Fig.~\ref{fig5} shows good
agreement of our analytical approach with numerical solutions of the NLS equation.

\section{Conservation of wave action}

The example of the preceding section demonstrates that though the packet undergoes drastic
transformations during its propagation (in particular, its amplitude becomes extremely small
in the region between the two large scale flows), nevertheless the value of the constant $q$
in Eq.~(\ref{eq31g}) remains the same at all stages of evolution. So one can predict the value of
the carrier wave number $k$ at any moment of time.

Equation (\ref{eq31g}) does not depend on the amplitude of the packet. Therefore it would be
desirable to have another conservation law which could provide some information about the
packet's amplitude. As is known, the packet's energy is not conserved in situations with
time-dependent external parameters of the large scale flow. Instead, as was shown by
Whitham \cite{whitham-65.2,whitham-65}, the wave action is preserved provided the external
parameters change slowly enough. Here we want to demonstrate that the packet's wave action remains
constant in spite of drastic packet's transformations shown the preceding section.

\begin{figure}[t]
\begin{center}
	\includegraphics[width = 6.5cm,height = 6.5cm]{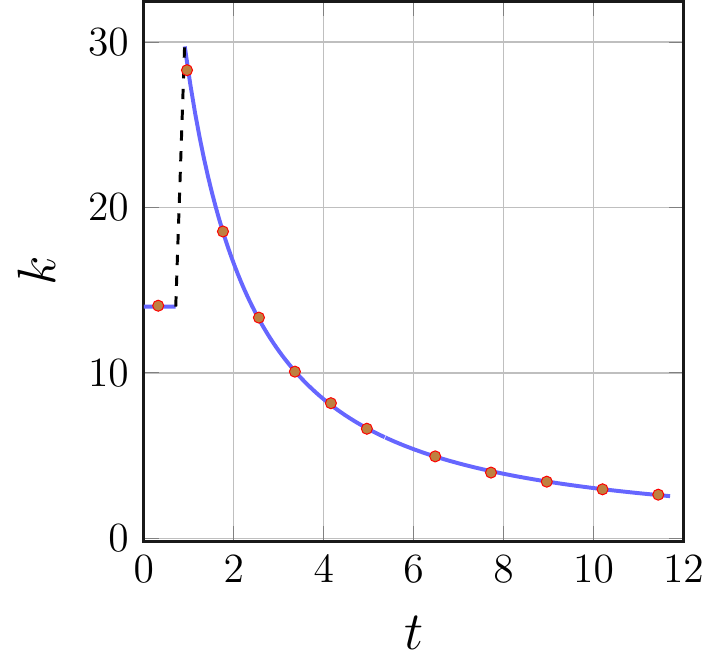}
\caption{Blue line denotes analytical prediction for the time dependence of the carrier wave number.
Circles correspond to numerical solutions of the NLS equation.
Black dashed line denotes transition between two expanding BEC clouds.
 }
\label{fig6}
\end{center}
\end{figure}

At first we shall derive by the Whitham method \cite{whitham-65.2,whitham-65} the expression
for the wave action in case of gNLS equation (\ref{eq26}). Linearization of the hydrodynamic-like
equations (\ref{eq28}) with respect to small packet's variables $\rho'=\rho-\overline{\rho},
u'=u-\overline{u}$, where $\overline{\rho}, \overline{u}$ obey the dispersionless equations
(\ref{eq7}), yields
\begin{equation}\label{eq44}
\rho'_t + \overline{u}\rho'_x + \overline{\rho}\phi_{xx} = 0, \qquad
\phi_t + \overline{u}\phi_x+\frac{c^2}{\overline{\rho}}\rho' -\frac{\rho'_{xx}}{4\overline{\rho}}= 0,
\end{equation}
where we have also introduced the potential $\phi$ of the packet's contribution to the
flow velocity $u'=\phi_x$. These equations can be formally derived from the principle
of minimal action with the Lagrangian density
\begin{equation}\label{eq45}
\mathop{\mathcal{L}} = \phi_t\rho' + \overline{u}\rho' \phi_x + \frac{c^2}{2\overline{\rho}}\rho'^2+
\frac{\overline{\rho}}{2}\phi_x^2 + \frac{(\rho'_x)^2}{8\overline{\rho}},
\end{equation}
where  $\overline{\rho}, \overline{u}$ are known functions and they should not be varied in our
variational problem (actually they can be considered constant at the packet's length scale).
Now we define the packet's envelope variables by the formulas
\begin{equation}\label{eq45b}
\rho' = A(x,t)\cos{\theta(x,t)}, \quad 
\phi = \frac{B(x,t)}{\theta_x}\sin{\theta(x,t) },
\end{equation}
where $\theta(x,t)$ is the high-frequency phase with slowly changing wave vector $k=\theta_x$
and frequency $\om=-\theta_t$. Substitution of Eqs.~(\ref{eq45b}) into first Eq.~(\ref{eq44})
gives the relationship
\begin{equation*}
B= \frac{A}{\overline{\rho}k}\big( \om - \overline{u}k \big).
\end{equation*}
Consequently, the Lagrangian density becomes proportional to $A^2$ and averaging of
$\mathcal{L}$ over fast oscillations at the packet's wavelength scale yields
\begin{equation}\label{eq46}
\overline{\mathop{\mathcal{L}}} = \frac{A^2}{\overline{\rho}}
\left\{c^2+\frac{1}{4}\theta_x^2  -
\left(\frac{\theta_t}{\theta_x} + \overline{u}  \right)^2  \right\},
\end{equation}
where we have omitted an inessential numerical factor 1/4.

\begin{figure}[t]
\begin{center}
	\includegraphics[width = 6.5cm,height = 6.5cm]{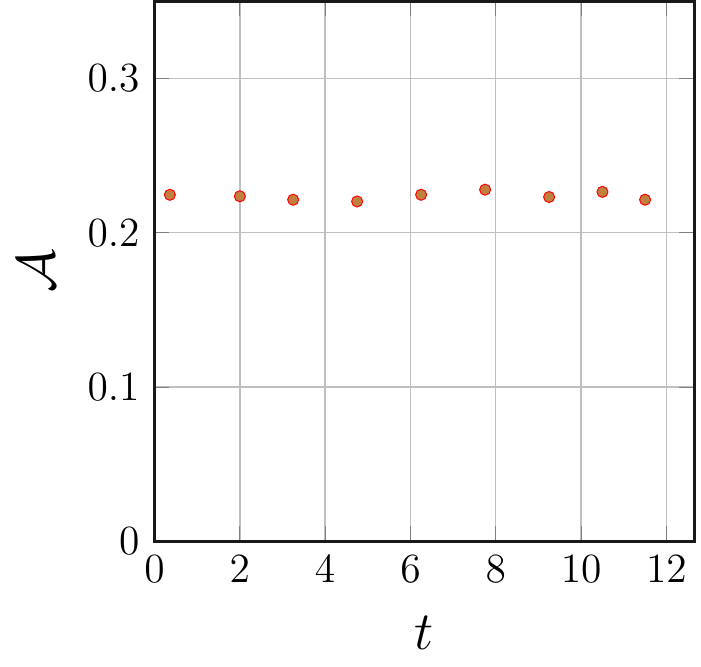}
\caption{Dependence of the packet's wave action (\ref{A})
during propagation of the packet along the path shown in Fig.~\ref{fig5}.
 }
\label{fig7}
\end{center}
\end{figure}

According to Whitham,
equations for the envelope variables are to be obtained from the variational problem for
the averaged action
\begin{equation}\label{eq47}
  S=\int\overline{\mathop{\mathcal{L}}}\,dxdt.
\end{equation}
Variation of $A$ gives
\begin{equation}\label{eq48}
  c^2+\frac{k^2}{4}=\left(\frac{\om}{k}-\overline{u}\right)^2
\end{equation}
and this equation reproduces the dispersion relation (\ref{eq29}). Variation of $\theta$
leads to the conservation law
\begin{equation}\label{eq49}
  \frac{\prt}{\prt t}\frac{\prt\mathop{\mathcal{L}}}{\prt\theta_t}+
  \frac{\prt}{\prt x}\frac{\prt\mathop{\mathcal{L}}}{\prt\theta_x}=0,
\end{equation}
where
\begin{equation}\label{eq50}
  \begin{split}
  & \frac{\prt\mathop{\mathcal{L}}}{\prt\theta_t}=
  \frac{2A^2}{\overline{\rho}k}\left(\frac{\om}{k}-\overline{u}\right)=
  \frac{A^2}{\overline{\rho}k}\sqrt{k^2+4c^2}\equiv w,\\
  & \frac{\prt\mathop{\mathcal{L}}}{\prt\theta_x}=
   \frac{A^2}{\overline{\rho}k}\sqrt{k^2+4c^2}\left(\overline{u}+\frac{k^2+2c^2}{\sqrt{k^2+4c^2}}\right)
   \equiv v_gw.
  \end{split}
\end{equation}
Thus, we arrive at the wave action conservation law
\begin{equation}\label{eq51}
  \frac{\prt w}{\prt t}+\frac{\prt(v_gw)}{\prt x}=0,
\end{equation}
where $w$ is defined in Eq.~(\ref{eq50}). If the packet is localized, then its total action
\begin{equation}\label{A}
\mathcal{A} = \int\limits_{- \infty}^{+ \infty}\frac{A^2}{\overline{\rho}k}\sqrt{4c^2+k^2}\,dx
\int\limits_{- \infty}^{+ \infty} \frac{A^2}{\overline{\rho}k}|q-\overline{u}|dx
\end{equation}
should be preserved during propagation through a slowly changing background (we have used
Eq.~(\ref{eq31g}) in the last expression for the wave action for the case $\ga=2$).
The wave action $\mathcal{A}$
depends on the amplitude $A$ of the packet and can be used for estimates of $A$, if the
packet's size can be measured independently.

We have checked conservation of the wave action (\ref{A}) during propagation of the
wave packet along the path discussed in the preceding section (see Figs.~\ref{fig4}-\ref{fig6}).
As one can see in Fig.~\ref{fig7}, the wave action calculated numerically keeps the constant
value in spite of quite steep turns in the packet's trajectory.

\section{Stationary background flow}

Our approach assumes that the background flow is described by Eqs.~(\ref{eq7})
which do not include external forces: if we add a term $-U_x$ corresponding to the
external potential $U(x)$ acting on the gas, then we cannot derive Eqs.~(\ref{eq20})
or (\ref{eq25}) anymore, so $k$ cannot be a function only of $\rho$ and $u$.
However, in an important particular case of a stationary background flow described by
the equations
\begin{equation}\label{eq52}
  (\rho u)_x=0, \qquad uu_x+\frac{c^2}{\rho}\rho_x=-U_x,
\end{equation}
Eqs.~(\ref{eq20}) or (\ref{eq25}) are not necessary for finding a path of
the high-frequency wave packet. Indeed, in this case the distributions $\rho=\rho(x)$,
$u=u(x)$ do not depend on time, consequently the frequency $\om=\om(k,\rho,u)$
obtained from linearized equations does not depend on time either, in agreement
with Hamilton equations (\ref{eq19}), i.e., $\om$ is their `energy integral'. This means
that the dependence $k=k(x)$ can be found from the frequency conservation law
\begin{equation}\label{eq53}
  \om(k,\rho(x),u(x))=\om_0=\mathrm{const},
\end{equation}
and then the packet's path can be found by integration of the equation
\begin{equation}\label{eq54}
  \frac{dx}{dt}=\left.\frac{\prt}{\prt k}\om(k,\rho(x),u(x))\right|_{k=k(x)}
\end{equation}
with a given initial condition.

\begin{figure}[t]
\begin{center}
	\includegraphics[width = 6.5cm,height = 6.5cm]{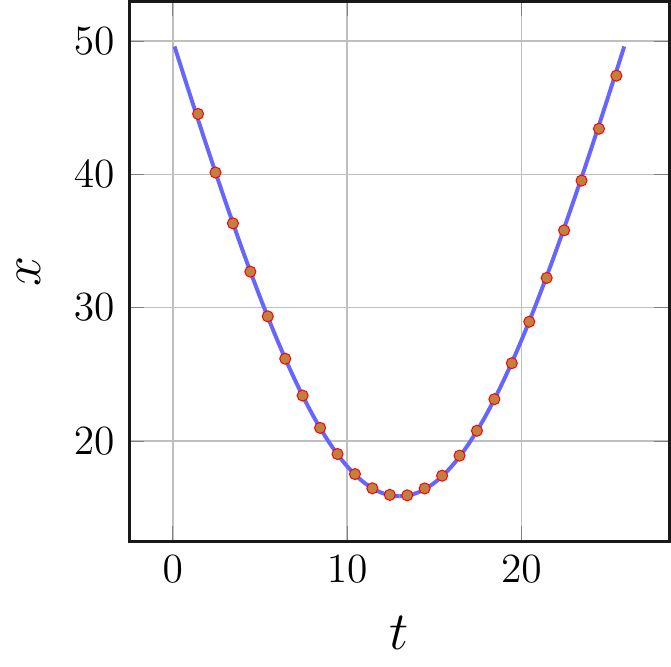}
\caption{Wave packet's path along a stationary flow.
 }
\label{fig8}
\end{center}
\end{figure}

We illustrate here this situation by a simple example for the NLS equation case with
$c^2=\rho$ ($\ga=2$). Equations (\ref{eq52}) can be integrated once to give
\begin{equation}\label{eq55}
  \rho u=\rho_0u_0,\qquad \frac12u^2+\rho+U=\frac12u_0^2+\rho_0,
\end{equation}
where $\rho_0, u_0$ are the flow variables far from the obstacle whose action is described
by the potential $U(x)$, $U\to0$ as $|x|\to\infty$. Elimination of $\rho=\rho_0u_0/u$
yields the equation
\begin{equation}\label{eq56}
  \frac12\left(u_0^2-u^2(x)\right)+\rho_0\left(1-\frac{u_0}{u(x)}\right)=U(x)
\end{equation}
which defines the function $u(x)$ for all values of $x$, $-\infty<x<\infty$, if the
maximal value $U_m>0$ of the potential $U(x)$ satisfies the condition (see \cite{legk-09})
\begin{equation}\label{eq57}
  U_m\leq\frac12u_0^2-\frac32(\rho_0u_0)^{2/3}+\rho_0.
\end{equation}
This inequality becomes equality for two values of $u_0$, $u_0=u_{\pm}$, $u_-<u_+$, so the
smooth solutions of Eq.~(\ref{eq56}) exist for subcritical $u_0<u_-$ and supercritical
$u_0>u_+$ flows.

\begin{figure}[t]
\begin{center}
	\includegraphics[width = 6.5cm,height = 6.5cm]{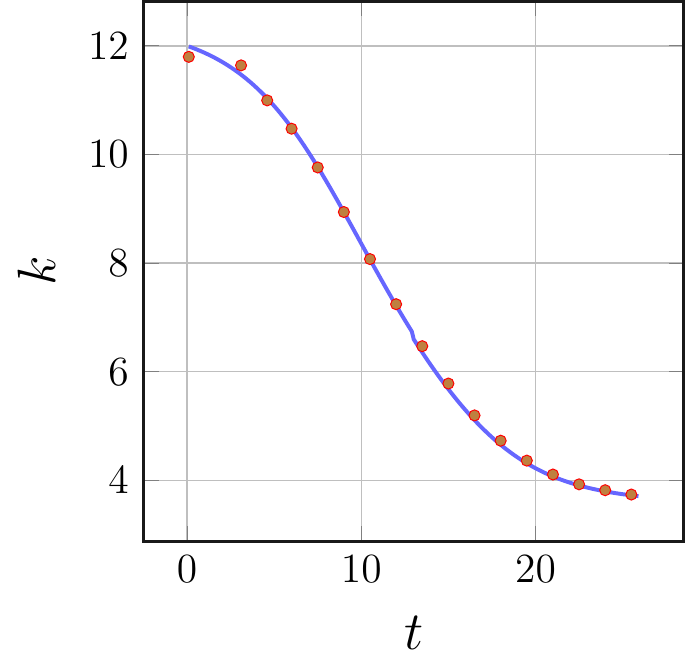}
\caption{Dependence of the carrier wave number during packet's propagation along
a stationary flow.
 }
\label{fig9}
\end{center}
\end{figure}

Let the potential be given by the expression
\begin{equation}\label{eq58}
  U(x)=\frac{U_m}{\cosh(x/\sigma)}
\end{equation}
with $U_m=15, \sigma=15$ and the flow at infinity have the values of the parameters
$\rho_0=1, u_0=8$. We launch the high-frequency wave packet at $x_0=50$ with the initial
carrier wave number $k_0=12$. The functions $u(x)$ and $\rho(x)=\rho_0u_0/u(x)$ are found
by numerical solution of Eq.~(\ref{eq56}) and then the function $k(x)$ is defined by
Eq.~(\ref{eq53}). The path $x=x(t)$ obtained by solving Eq.~(\ref{eq54}) is shown
in Fig.~\ref{fig8} by a solid line and it agrees very well with the exact numerical solution
of the NLS equation whose results are shown by dots. When $x=x(t)$ is known, then the
time dependence of the carrier wave number $k=k(x(t))$ is known, too; it is shown
in Fig.~\ref{fig9} and also agrees with the numerical solution of the NLS equation.
The final value $k=k_f$ at $t\to\infty$ can be predicted from the conservation law
(\ref{eq53}),
\begin{equation}\label{eq59}
  k_f\left(u_0-\sqrt{\rho_0+\frac{k_f^2}{4}}\right)=
  k_0\left(u_0-\sqrt{\rho_0+\frac{k_0^2}{4}}\right),
\end{equation}
where the sign before the square root is chosen in such a way that the initial group velocity
(\ref{eq17}) is negative. Then we obtain $k_f\approx3.988$, that is the carrier wave
vector remains positive during propagation along this stationary flow, but the group velocity
changes sign at the turning point, so the packet changes direction of its propagation
(see Fig.~\ref{fig8}).

\section{Conclusion}

In this paper, we have extended the approach proposed in Ref.~\cite{MU} for description of propagation of
high-frequency wave packets along a large scale simple wave to motion of such wave packets along
large-scale background pulses described by general solutions of hydrodynamic type systems for
two variables $\rho$ and $u$. Due to large difference of scales inherent in our task, the wave
packet's motion is separated from the background pulse evolution. Consequently, the background wave
variables $\rho,u$ evolve according to dispersionless
equations, whereas propagation of the wave packet is governed by the Hamilton equations with the
carrier wave frequency $\om$ and wave vector $k$ playing the roles of the Hamiltonian and canonical momentum,
respectively. We have formulated conditions under which fulfillment the wave vector $k$ is a
function of two variables $\rho,u$. Typically, these conditions are fulfilled for
high-frequency packets in the limit of large $k$. In this case, the combined system
of Hamilton equations and dispersionless hydrodynamic equations can be reduced to equations for
$k$ as a function of the background wave variables $\rho,u$. Asymptotic solution of these equations
for the gNLS equation case has a quite simple general form (\ref{eq31f}) and it allows
one to find the packet's path and the dependence $k(t)$ of the carrier wave vector on time.
Application of this theory to concrete examples  demonstrated its
good agreement with the exact numerical solution of the full system.

\begin{acknowledgments}
We thank S.~K.~Ivanov for useful discussions.
This research was funded by the research project FFUU-2021-0003 of the Institute of Spectroscopy of
the Russian Academy of Sciences (sections I-IV) and by the Foundation for
the Advancement of Theoretical Physics and Mathematics
``BASIS'' (sections V-VII).
\end{acknowledgments}

\section*{Data availability}
The data that support the findings of this study are available from the corresponding author upon reasonable request.

\end{document}